%% file: main.tex
\documentclass[sigplan,nonacm]{acmart}
\copyrightyear{2026}
\acmYear{2026}
\setcopyright{cc}
\setcctype{by}
\acmConference[ASPLOS '26] {Proceedings of the 30th ACM International Conference on Architectural Support for Programming Languages and Operating Systems, Volume 2}{March 22--26, 2026}{Pittsburgh, PA, USA.}
\acmBooktitle{Proceedings of the 30th ACM International Conference on Architectural Support for Programming Languages and Operating Systems, Volume 2 (ASPLOS '26), March 22--26, 2026, Pittsburgh, PA, USA}
\acmISBN{979-8-4007-2359-9/2026/03}
\acmDOI{10.1145/3779212.3790123} 
\acmSubmissionID{V2fp0576}

\usepackage{amsmath,amsfonts}
\usepackage{physics}
\usepackage{algorithmicx}
\usepackage{algorithm}
\usepackage{graphicx}
\usepackage{subcaption}
\usepackage{textcomp}
\usepackage[dvipsnames]{xcolor}
\usepackage{diagbox}
\usepackage{multirow}
\usepackage{multicol}
\usepackage{makecell}
\usepackage{enumitem}
\usepackage{hyperref}
\usepackage{adjustbox}
\usepackage[usestackEOL]{stackengine}
\usepackage[noend]{algpseudocode}
\usepackage{balance}
\usepackage[T1]{fontenc}
\hypersetup{}

\newcommand{\revdelete}[1]{}
\newcommand{\revadd}[1]{#1}

\newcommand{\myCompilerName}{AlphaSyndrome} 
\newcommand{\myCompilerNameSpace}{AlphaSyndrome }

\begin{document}

\author{Yuhao Liu}
\orcid{0009-0005-2822-0448}
\affiliation{%
  \institution{University of Pennsylvania}
  \city{Philadelphia}
  \country{United States}
}
\email{liuyuhao@seas.upenn.edu}

\author{Shuohao Ping}
\orcid{0009-0008-8723-3208}
\affiliation{%
  \institution{University of Pennsylvania}
  \city{Philadelphia}
  \country{United States}
}
\email{sp1831@seas.upenn.edu}

\author{Junyu Zhou}
\orcid{0009-0007-5564-8401}
\affiliation{%
  \institution{University of Pennsylvania}
  \city{Philadelphia}
  \country{United States}
}
\email{junyuzh@seas.upenn.edu}

\author{Ethan Decker}
\orcid{0009-0008-0567-9143}
\affiliation{%
  \institution{University of Pennsylvania}
  \city{Philadelphia}
  \country{United States}
}
\email{ecd5249@seas.upenn.edu}

\author{Justin Kalloor}
\orcid{0000-0001-6749-5501}
\affiliation{%
  \institution{University of California, Berkeley}
  \city{Berkeley}
  \country{United States}
}
\email{justin.kalloor@gmail.com}

\author{Mathias Weiden}
\orcid{0000-0003-4384-1099}
\affiliation{%
  \institution{University of California, Berkeley}
  \city{Berkeley}
  \country{United States}
}
\email{mtweiden@berkeley.edu}

\author{Kean Chen}
\orcid{0000-0002-0772-6635}
\affiliation{%
  \institution{University of Pennsylvania}
  \city{Philadelphia}
  \country{United States}
}
\email{keanchen@seas.upenn.edu}

\author{Yunong Shi}
\orcid{0000-0002-0824-6107}
\affiliation{%
  \institution{Amazon Quantum Technologies}
  \city{Pasadena}
  \country{United States}
}
\affiliation{%
  \institution{University of Michigan, Ann Arbor}
  \city{Ann Arbor}
  \country{United States}
}
\email{synisnot@gmail.com}

\author{Ali Javadi-Abhari}
\orcid{0000-0002-8022-2695}
\affiliation{%
  \institution{IBM Research}
  \city{Yorktown Heights}
  \country{United States}
}
\email{ali.javadi.abhari@gmail.com}

\author{Costin Iancu}
\orcid{0000-0001-7845-2427}
\affiliation{%
  \institution{Lawrence Berkeley National Laboratory}
  \city{Berkeley}
  \country{United States}
}
\email{cciancu@lbl.gov}

\author{Gushu Li}
\orcid{0000-0002-6233-0334}
\affiliation{%
  \institution{University of Pennsylvania}
  \city{Philadelphia}
  \country{United States}
}
\email{gushuli@seas.upenn.edu}

\renewcommand{\shortauthors}{Yuhao Liu et al.}

\ccsdesc[500]{Computer systems organization~Quantum computing}
\ccsdesc[500]{Hardware~Quantum error correction and fault tolerance}
\ccsdesc[500]{Software and its engineering~Compilers}

\keywords{Quantum Computing; Quantum Error Correction; Syndrome Measurement}

\received{21 August 2025}
\received[revised]{8 January 2026}
\received[accepted]{16 January 2026}

\title{\myCompilerName: Tackling the Syndrome Measurement Circuit Scheduling Problem for QEC Codes}

\begin{abstract}
Quantum error correction (QEC) is essential for scalable quantum computing, yet repeated syndrome-measurement cycles dominate its spacetime and hardware cost. Although stabilizers commute and admit many valid execution orders, different schedules induce distinct error-propagation paths under realistic noise, leading to large variations in logical error rate. Outside of surface codes, effective syndrome-measurement scheduling remains largely unexplored.
We present AlphaSyndrome, an automated synthesis framework for scheduling syndrome-measurement circuits in general commuting-stabilizer codes under minimal assumptions: mutually commuting stabilizers and a heuristic decoder. AlphaSyndrome formulates scheduling as an optimization problem that shapes error propagation to (i) avoid patterns close to logical operators and (ii) remain within the decoder's correctable region. The framework uses Monte Carlo Tree Search (MCTS) to explore ordering and parallelism, guided by code structure and decoder feedback.
Across diverse code families, sizes, and decoders, AlphaSyndrome reduces logical error rates by 80.6\% on average (up to 96.2\%) relative to depth-optimal baselines, matches Google's hand-crafted surface-code schedules, and outperforms IBM's schedule for the Bivariate Bicycle code.
\end{abstract}

\maketitle

\input{tex/1-introduction}
\input{tex/2-background}

\input{tex/3-syndrome}
\input{tex/4-method}

\input{tex/5-evaluation}
\input{tex/6-conclusion}
\input{tex/7-acknowledgement}

\bibliographystyle{ACM-Reference-Format}
\balance
\bibliography{refs}

\input{tex/99-appendix}

\end{document}

%% file: tex/1-introduction.tex
\section{Introduction}\label{sec:introduction}

Quantum error correction (QEC)~\cite{nielsen2010quantum,gottesman1997stabilizer} is the mechanism that makes scalable quantum computation physically plausible. Physical qubits inevitably suffer from decoherence, control imperfections, leakage, and measurement noise, so raw error rates compound quickly with circuit depth. QEC counters this by encoding a logical qubit into a structured, redundant subspace and continually detecting and removing errors without disturbing the encoded information. When paired with fault‑tolerant implementations of logical operations~\cite{gottesman2024surviving}, QEC allows logical error rates to be suppressed — often exponentially with code distance — provided the underlying physical error rates are below threshold and sufficient overhead is invested. As a result, any architecture aiming at reliable, large‑scale quantum algorithms must devote the bulk of its spacetime and hardware resources to QEC.
Recently, early QEC experiments have been demonstrated on various hardware platforms~\cite{cory1998experimental,lukin2023logical,acharya2024qecsurface,acharya2022suppressing,reichardt2025logical,wu2022erasure,livingston2022experimental}. 

In the widely adopted stabilizer‑based QEC, protection is realized through repeated syndrome‑measurement cycles. Each cycle measures a mutually commuting set of stabilizer generators whose outcomes (the syndrome) reveal the presence and type of physical errors while leaving the logical state unchanged. A standard, hardware‑agnostic realization uses ancilla‑mediated measurement: prepare an ancilla; apply a sequence of two‑qubit Pauli checks that entangle the ancilla with the data qubits in the support of a stabilizer (implemented with CNOT/CZ gates and simple single‑qubit pre/post rotations for X, Y, or Z checks); measure and reset the ancilla; and stream the results to a classical decoder. These subroutines dominate the spacetime footprint of an error‑corrected program and are executed at high repetition rates across many rounds. 

The design space for implementing the syndrome measurement circuit of one QEC code is extremely large.
Different from a normal quantum circuit, where the quantum gates do not commute most of the time, there exists great commutativity in the syndrome measurement circuit.
For most QEC codes, all the stabilizers are constructed to commute with each other so that their corresponding syndrome measurement circuits also mostly commute.
Moreover, those Pauli checks between the data qubits and the ancilla qubit inside one syndrome measurement circuit also mostly commute. 
If there were no errors in the physical circuit, many orders of executing the Pauli checks and the syndrome measurements in the large design space would be essentially the same -- doing nothing to the logical state.

\begin{figure}[t]
    \centering
    \includegraphics[width=\linewidth]{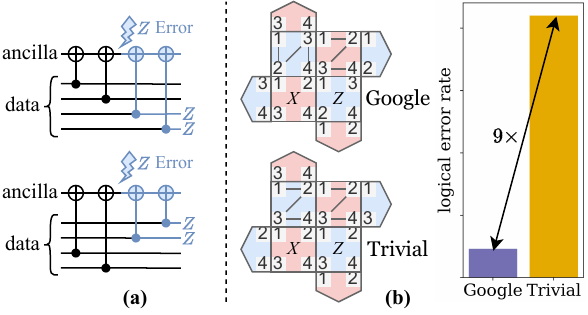}
    \caption{(a) Two different schedules of the same $ZZZZ$ syndrome measurement circuit. (b) The logical error rates of two different syndrome measurement circuit schedules. The numbers in the surface code patches represent the time steps when the corresponding Pauli check between the corresponding data qubit and ancilla qubit is executed.}
   \Description{..}
    \label{fig:motivation}
\end{figure}

However, unlike the ideal case, the inevitable physical error makes different orderings of the syndrome measurement circuits behave significantly differently.
The reason is that the errors may propagate from one qubit to another qubit via the two-qubit gates in the syndrome measurement circuits.
For example, an error happens in the middle of a 4-qubit syndrome measurement on the ancilla qubit can propagate to two different data qubits under the two different schedules shown in Figure~\ref{fig:motivation} (a).
The order of the gates will affect the direction in which errors may propagate and will significantly impact the final logical error rate. Figure~\ref{fig:motivation} (b) shows the logical error rate of syndrome measurement for a distance-3 rotated surface code. The logical error rate under a trivial syndrome measurement circuit schedule can be 9$\times$ worse than that under the good schedule designed by Google~\cite{acharya2024qecsurface}.

It turns out that, to the best of our knowledge, the schedule of the syndrome measurement circuit is mostly \textit{hand-crafted} and not well-optimized.
Common approaches include the trivial qubit index ordering~\cite{zhao2022realization,ryan2021realization} as well as the lowest depth scheduling~\cite{kang2025quits,menon2025magic}, which aims to minimize the effect of decoherence.
Google has carefully developed good manual designs~\cite{acharya2024qecsurface,horsman2012surface} for the rotated surface code by leveraging its well-known background topology and the translational invariance property.
Yet, most QEC codes do not have such good properties, and the syndrome measurement circuit schedule for the surface code cannot be directly generalized to other codes.
Moreover, the \textit{non-uniformity in the device noise} and \textit{the approximation in the decoding algorithm}, both of which are inevitable in practice, are not yet considered.

In this paper, we address this open problem and propose \myCompilerName, a \revdelete{compilation}\revadd{synthesis} framework that can automatically \revdelete{compile}\revadd{synthesize} and optimize the syndrome measurement circuit schedule for QEC codes with very minimal and widely held assumptions -- all stabilizers commute with each other, and the decoder is heuristic.
\textbf{First}, we formulate the problem of scheduling the syndrome measurement circuit in the general case and made a key observation that the error propagation should be optimized towards two objectives: the error should propagate to an error pattern that is 1) \textit{not close to a logical operation}, and 2) \textit{within the correct decoding capability of the heuristic decoder}. Optimizing towards these two objectives will lower the logical error rate and improve the performance of the QEC codes.
\textbf{Second}, based on this key observation, \myCompilerNameSpace employs a Monte-Carlo-Tree Search-based framework to learn to optimize the scheduling of the gates in syndrome measurement circuits.
Such a data-driven approach can learn from the code structures and decoding algorithms under a given noise model, considering all the factors to schedule the syndrome measurement circuit with much lower logical error. 

We evaluated \myCompilerNameSpace on various QEC codes of different code families, sizes, and decoders. 
The results show that optimizing for the lowest depth is \textit{not} necessarily a good design choice, and \myCompilerNameSpace can outperform lowest depth baseline schedules with a logical error rate reduction of 80.6\% on average (up to 96.2\%).
In particular, \myCompilerNameSpace can match the performance of the known good schedule on rotated surface codes designed by Google, and outperform IBM's schedule on the bivariate bicycle code.

Our major contributions can be summarized as follows:
\begin{enumerate}
    \item We propose \myCompilerName, a \revdelete{compiler}\revadd{synthesis} framework to automatically optimize the schedule of syndrome measurement circuits for QEC codes based on the selected decoder and the noise model.
    \item We made the key observation that lowest depth schedules can be far from performance-optimal implementations, and summarize two optimization objectives with respect to the code structure and the decoder, respectively.
    \item We design a Monte-Carlo Tree Search (MCTS) based synthesis framework to synthesize and optimize the syndrome measurement circuit for various QEC codes and corresponding decoders under a given noise model.
    \item Evaluation results show that our \revdelete{compiled}\revadd{synthesized} syndrome measurement circuit schedule can outperform the lowest depth schedules for various QEC codes of different families, sizes, and decoders. In particular, we can match the performance of Google's schedule for rotated surface codes and outperform IBM's schedule for the bivariate bicycle code. 
\end{enumerate}

%% file: tex/2-background.tex
\section{Background}

This section provides background knowledge about quantum computing, QEC, stabilizer formalism, and Monte Carlo tree search. We do not cover the quantum computing basics, and we recommend~\cite{nielsen2010quantum} for more details.

\subsection{Quantum Error Correction Basics}

In the real world, the environment can easily affect quantum information, and this fluctuation disrupts computation by causing quantum errors. \revadd{When transmitting or computing on a state $\ket{\psi}$, a set of errors $\{E_i\}$ could randomly apply to the state with probability $\{p_i\}$. The random operations represent quantum errors, and our goal is to detect a specific $E_i$ that has occurred to the state and correct it. Classical methods won't apply, as}\revdelete{Moreover,} the no-cloning theorem~\cite{nielsen2010quantum} forbids us from duplicating quantum information to confront error, \revadd{and measuring the state voids the internal information}. Thus, a dedicated quantum error correction system is devised to detect, correct, and counter quantum errors. \revadd{The system consists of two parts: a quantum error correction code (QECC) that encodes information redundantly, and an error correction process that recovers the polluted information.}

\subsubsection{Quantum Error Correction Code (QECC)}

A QECC $C$ \revdelete{is an encoding of}\revadd{encodes} $k$ logical qubits using \(n\) physical qubits. Mathematically, a $2^k$-D subspace of the $2^n$-D state space of $n$ physical qubits is used to represent the $2^k$-D Hilbert space of $k$ logical qubits. The subspace is called the \textit{code space}. 

The rest of the state space is used to detect errors by performing projection measurements onto it. This projective measurement process is called \textit{syndrome measurement}. \revadd{The projective measurement allows us to tell whether errors happened (leave the code space) without touching the exact state.} Subsequently, according to the measurement results, a classical decoder is applied to decide what errors occurred on the physical qubits. Inverse error gates are then applied and controlled by the decoder. The two processes together form the \textbf{error correction} procedure of a QECC.

\subsubsection{Stabilizer Code}

Almost all QECCs are \textbf{stabilizer codes}. A stabilizer code of $n$ physical qubits is defined by a \revdelete{set} \revadd{group} $\mathcal{G}$ \revdelete{of} \revadd{generated by} $r$ commuting and algebraic independent (cannot be expressed by the production of other elements) Pauli strings $\mathcal{G}=\langle S_1,\dots, S_r\rangle$. They form the \textit{stabilizer group} under multiplication, and elements in $\mathcal{G}$ are the stabilizers of the code. The code space is defined as the common $+1$ eigenspace of stabilizers, that is, spanned by $\{\ket{\psi}\ |\ \forall S_i\in\mathcal{G}, S_i\ket{\psi}=\ket{\psi}\}$. \revadd{Each Pauli string $S_i\in\mathcal{G}$ splits the Hilbert space by half, thus }the subspace is of dimension $2^n/2^r=2^{n-r}$ and encodes $k=n-r$ logical qubits.

For example, the rotated surface code is shown in Figure~\ref{fig:logicops} (a). It has 25 data qubits (dots) and 24 stabilizers (the \revadd{red and blue} colored squares in the middle and semicircles at the edges), encoding $25-24=1$ logical qubits. 
For each stabilizer, it is either $X$-type (red) or $Z$-type (blue), meaning that we have $X$ or $Z$ operators on the data qubits at the corners of the red or blue squares/semicircles, respectively.

Any quantum error $E$ that anticommutes with a stabilizer $S_i$ \revadd{($ES_i+S_iE=0$)} \revadd{causes the state to fall into the $-1$ eigenspace of $S_i$ and} can be detected by \revadd{projective} measuring $S_i$, yielding an outcome of $-1$ instead of $+1$. The whole error correction process involves measuring all stabilizers and interpreting the measurement outcome to determine the exact \revdelete{type}\revadd{combination} of errors that occurred. However, if an error commutes with all stabilizers and itself is not in $\mathcal{G}$, it can not be detected, causing a \textit{logical error}. The code distance of a stabilizer code is the minimal weight $d$ of a Pauli string that commutes with all stabilizers but does not belong to the stabilizer group. It is considered to be the smallest error that the code cannot detect. 
The notation $[[n,k,d]]$ denotes a stabilizer code that encodes $k$ logical qubits with $n$ physical qubits and has a distance $d$.

\subsubsection{Logical Operation and Logical Error}

A \textbf{logical operation} on a QECC is a gate $U$ that operates on data qubits, mapping the code space $\mathcal{C}$ to itself. Similarly, if a set of errors on data qubits behaves the same as a logical operation, they become a \textbf{logical error}. They should commute with all stabilizers, which ensures the state stays in the $+1$ eigenspace of stabilizers after $U$. On the logical level, $U$ performs a specific gate operation $U_L$ over the encoded logical qubits. The concrete semantics correspondence between $U$ and $U_L$ usually depends on the encoding isomorphism between the code space and the logical space $\mathcal{H}^{2^k}$, and the code structure.

\begin{figure}[t]
    \centering
    \includegraphics[width=0.9\linewidth]{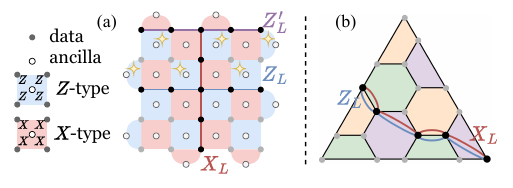}
    \caption{Example of logical $X$ and $Z$ operations on (a) $[[25,1,5]]$ rotated surface code and (b) $[[19,1,5]]$ hexagonal color code, each polygon is an $X$ and $Z$ stabilizer. Two $Z_L$s on the rotated surface code are equivalent by \revadd{multipling} the starred stabilizers.}
    \Description{..}
    \label{fig:logicops}
\end{figure}

In \revdelete{one}\revadd{a particular} QECC, one logical operation would have \textit{many different equivalent forms of realization}. That is, multiplying any element $S\in\mathcal{G}$ in the code's stabilizer group to the logical operation $U$ will yield another logical operation $U'=SU$ performing the same logical gate:  $\forall\ket{\psi}\in\mathcal{C},\ U'\ket{\psi}=SU\ket{\psi}=US\ket{\psi}=U\ket{\psi}$.
For example, on the rotated surface codes in Figure~\ref{fig:logicops} (a), we usually pick the vertical multi-qubit $X$ Pauli string (red) and horizontal multi-qubit $Z$ Pauli string (blue) as the logical $X_L$ and $Z_L$ gate. The two Pauli string anticommute just like the single-qubit $X$ and $Z$ gate, and it is easy to verify they commute with all stabilizers. Another $Z_L'$ (purple) is equivalent to $Z_L$ by multiplying the starred stabilizers. 
Actually, in the surface code patch, any chain of physical single-qubit $Z$ gates connecting the left and right edges realizes the same logical $Z_L$ gate. 
Similarly, any chain of physical single-qubit $X$ gates connecting the top and bottom edges will realize the same logical $X_L$ gate. 

Note that the structure of logical operations on a rotated surface code is clear, but cannot be directly generalized to other QECCs.
For example, Figure~\ref{fig:logicops} (b) shows that the logical $Z_L$ and $X_L$ gates on the $[[19,1,5]]$ hexagonal color code can be realized as a chain of physical $Z$ and $X$ gates, respectively.
For general QECC, the equivalent logical operations can be derived by multiplying elements in the stabilizer group, but we usually do \textit{not} have an intuitive interpretation of the logical operation structures like those of surface code.

\subsection{Syndrome Measurement and Decoding}\label{sec:syndromeanddecode}
\begin{figure}[t]
    \centering
    \includegraphics[width=0.85\linewidth]{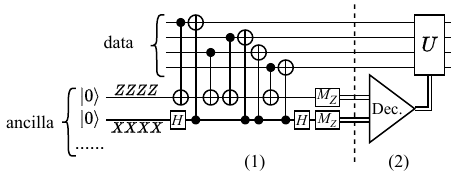}
    \caption{Error correction process of a QECC. (1) syndrome measurement, and (2) decoder and correction.}
    \Description{Error correction process of a QECC. (1) syndrome measurement, and (2) decoder and correction.}
    \label{fig:errorcorrection}
\end{figure}

QECC protects the quantum information by repeatedly executing the error correction process throughout the computation. As shown in Figure~\ref{fig:errorcorrection}, the error correction process of a stabilizer code contains two stages: (1) detect error syndromes through a \textbf{syndrome measurement} circuit, and (2) apply appropriate error correction \revadd{gates} through a classical procedure called \textbf{decoder}\revdelete{and controlled gates}. The syndrome measurement circuit measures all the stabilizers of the QECC using ancilla qubits, and the decoder~\cite{higgott2025blossom,roffe2020bposd,berlekamap1978on,lovasz1972factorization,liyanage2024fpga,liyanage2023scalable,overwater2022neural,varbanov2023neural} takes the syndrome measurement results as input and determines the type of errors that occurred on each data qubit, then applies the inverse as corrections.

\begin{figure}[ht]
    \centering
    \includegraphics[width=0.7\linewidth]{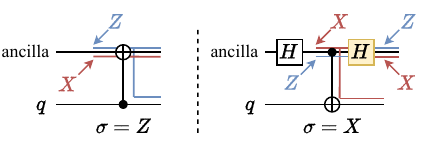}
    \caption{Two different Pauli checks: $Z$ check and $X$ check, and how error propagation during a Pauli-$X\text{ or }Z$ check. Notice the Hadamard gate (yellow) changes an error between being $X$ and $Z$. }
    \Description{..}
    \label{fig:hookpropagation}
\end{figure}

In syndrome measurements, we need to measure all the stabilizers by performing (multi-)Pauli checks on data qubits using ancilla qubits. For each stabilizer $S=\sigma_n\dots\sigma_1$ ($\sigma_i\in\{I,X,Z\}$), we introduce one ancilla qubit $q_{ancilla}$ for the measurement to collect result by preparing it into $\ket{0}$. Iterate through $S$, for $\sigma_i\neq I$, a short circuit is applied between the data qubit $q_i$ and the ancilla qubit $q_{ancilla}$ to perform a partial stabilizer measurement, as shown in Figure~\ref{fig:hookpropagation}. We will call it a \textit{partial Pauli-$\sigma_i$ check}. Finally, we perform a $Z$-basis measurement on the ancilla qubit $q_{ancilla}$.


The order to measure all stabilizers, particularly all Pauli checks, is not unique. Under an ideal error-free scenario, as long as the anticommutation relationship is satisfied~\cite{beverland2021cost,geher2024tangling} on the overlapped data qubits for different stabilizers, either order yields an identical outcome, and a trivial lexical or lowest depth order is often implemented in experiments.

\subsection{Monte Carlo Tree Search}

Monte Carlo tree search (MCTS) is a heuristic search algorithm. It operates by iteratively building a search tree, exploring promising moves more deeply while maintaining a balance between the exploration of new states and the exploitation of already known good states. 

The optimization problem MCTS operates on is formulated into several key concepts:
\begin{enumerate}
    \item \textbf{States}: the feasible region of the problem can be captured by finite and discrete states $R=\{S\}$.
    \item \textbf{Moves}: a move $m$ transforms a state $S$ into another state $S'$: $S\xrightarrow{m}S'$. Given $S$, its \textit{possible moves} are $mov(S)=\{m\ |\ \exists S'\in R,\ S\xrightarrow{m}S'\}$.
    \item \textbf{Termination}: a state $S$ is a termination state if it has no \textit{possible moves}. That is, $mov(S)=\emptyset$.
    \item \textbf{Evaluation}: a evaluation function $eva(S)$ will give a score $E$ to $S$ if $S$ is a termination state. The higher $E$, the better $S$.
\end{enumerate}
Starting from an initial state $S_{0}$, the ultimate goal is to find a path $S_{0}\xrightarrow{m_0}S_1\xrightarrow{m_2}\dots\xrightarrow{m_k-1}S_{k}$ towards a termination state $S_k$ such that it has the best score. Finding the entire path directly is challenging, and our optimization goal is to determine the next move $m_0$ that most likely leads to the best final state.

MCTS builds a search tree starting from the initial state $S_0$. In the search tree, each node carries a possible state of the search and two fields: its accumulated evaluation score $E$ and the number of visits $n$. The expectation score $\hat{E}$ of a state is defined as $E/n$. Each edge is a valid move. The tree is constructed with four phases, as shown in Figure~\ref{fig:mcts}:
\begin{figure}[t]
    \centering
    \includegraphics[width=0.9\linewidth]{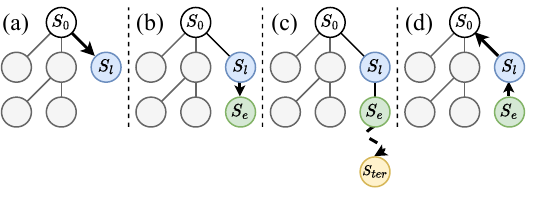}
    \caption{Four phases of MCTS. (a) Selection. (b) Expansion. (c) Simulation. (d) Backpropagation.}
    \Description{Four phases of MCTS}
    \label{fig:mcts}
\end{figure}

\begin{enumerate}[label=(\alph*)]
    \item \textbf{Selection:} Starting from the initial state $S_0$, the algorithm traverses downward the tree by selecting child nodes that maximize the "upper confidence bound applied to trees" (UCT)~\cite{kocsis2006bandit}. For the $i$th child:
    $UCT_i = \hat{E_i} + c\sqrt{\frac{\ln N}{n_i}}$
    where $\hat{E_i}$ is the expectation score of the $i$th child, $n_i$ is the number of times the child has been visited, $N$ is the total number of times the parent node has been visited, and $c$ is an exploration parameter, usually $\sqrt{2}$.

    \item \textbf{Expansion:} The \textbf{selection} is continuously applied downward to the tree until either a termination or a not fully explored (it has non-child next states) node $S_l$. If $S_l$ is a termination state, then define $S_e:=S_l$ and proceed to the next step. Otherwise, we create a new child $S_{new}$ of $S_l$ if $\exists m\in mov(S_l),S_l\xrightarrow{m}S_{new}$ and $S_{new}$ is not already $S_l$'s child. We set $S_e:=S_{new}$. $S_e$ is referred to as the "expanded node".

    \item \textbf{Simulation:} From $S_e$, a "roll-out" is performed to the termination. Specifically, we continuously select a random move starting from $S_e$ until we reach a termination state $S_{ter}$. An evaluation is applied to $eva(S_{ter})=E$.

    \item \textbf{Backpropagation:} The score $S$ is then propagated back up the tree from $S_e$ to the root $S_0$. Along the path, all nodes accumulate the score $E$ and increase their visit count by 1.
\end{enumerate}

A certain number of iterations will repeat the four phases, and the best next move $m_0$ is selected based on maximizing the expectation score $\hat{E}_i$ of the $S_0$'s $i$th child.

%% file: tex/3-syndrome.tex
\section{Problem Formulation and Analysis}\label{sec:syndromemeasurement}


The problem of scheduling the syndrome measurement circuit arises because the circuit itself is prone to errors.
Unlike the ideal error-free case, where different schedules yield precisely the same results, the inevitable errors in the syndrome measurement circuit make different schedules behave significantly differently.
In this section, we formulate this problem by analyzing the error propagation in the syndrome measurement and the design objectives we target.

\subsection{Errors Propagation in Syndrome Measurement}
We first analyze the error propagation in a single Pauli check.
We explain it in a common setting where the physical two-qubit gate between the data qubit and the ancilla qubit is CNOT for simplicity, and the discussion can be easily generalized to other two-qubit gates like CZ.
As shown in Figure~\ref{fig:hookpropagation}, the CNOT gate with the data qubit as the control and ancilla qubit as the target qubit implements the Pauli-Z check.
In this case, a $Z$ error in the ancilla qubit can propagate from the ancilla qubit to the data qubit, and an $X$ error will stay on the ancilla qubit.
In contrast, for the CNOT gate in the Pauli-X check circuit, the ancilla qubit is the control, and the data qubit is the target qubit.
An $X$ error can propagate from the ancilla qubit to the data qubit, and a $Z$ error will not.
In summary, a Pauli-$\sigma$ ($\sigma \in \{Z, X\}$) check will propagate a $\sigma$ error from the ancilla qubit to the data qubit.


\begin{figure}[t]
    \centering
    \includegraphics[width=0.95\linewidth]{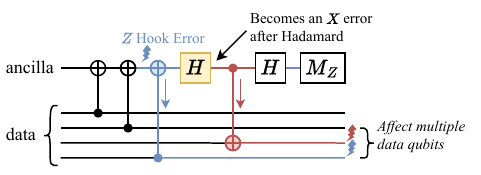}
    \caption{Syndrome measurement circuit of $ZZXZ$ stabilizer. A $Z$ hook error on the ancilla qubit propagates to multiple data qubits, causing $Z$ and $X$ errors.}
    \Description{A $Z$ hook error that propagates to multiple data qubits, causing $Z$ and $X$ errors.}
    \label{fig:hookmultiple}
\end{figure}

We then analyze the error propagation in a syndrome measurement circuit for a stabilizer with multiple Pauli checks.
An example of the syndrome measurement circuit for a stabilizer $ZZXZ$ is in Figure~\ref{fig:hookmultiple}.
As introduced in~\ref{sec:syndromeanddecode}, the order of implementing the four Pauli checks can be flexible. Suppose we first do the Pauli-$Z$ checks, followed by the last $X$ check. Figure~\ref{fig:hookmultiple} shows the circuit of such scheduling.
When a $Z$ error happens after the second check, this error will propagate to two data qubits through the remaining two Pauli checks, causing $Z$ errors on one data qubit and $X$ errors on one data qubit.
Such an error on the ancilla qubits is usually denoted as a \textit{hook error} in literature~\cite{dennis2002topological}.
A hook error is critical to the QEC performance because it can propagate to multiple data qubits and significantly increase the logical error rate.

A hook error makes the schedule of the syndrome measurement circuit matter, since the probability of being affected by the hook error for different data qubits can be affected by the order of executing the Pauli checks.
The hook error can happen anytime during the execution of the syndrome measurement circuit. It will only affect the data qubits whose Pauli checks happen after the hook error and will not affect the data qubits whose Pauli checks have completed.
As the error effect is equivalent up to a stabilizer, a hook error can at most affect the second half or the first half data qubits in the measurement sequence.
Therefore, a data qubit whose corresponding Pauli check is executed later would be more vulnerable since a hook error is more likely to affect it.
For the example in Figure~\ref{fig:hookmultiple}, an error on the second position affects three data qubits, but by multiplying a stabilizer, it is equivalent to having a $Z$ error on the first data qubit.

In summary, different scheduling of the syndrome measurement circuit will affect how errors are likely to occur on different data qubits.
Naturally, to reduce the overall logical error rate, our research question is \textit{how we should schedule the syndrome measurement circuit so that the errors are more likely to propagate to a `good' pattern that is correctable without causing a logical error}.
In the rest of this section, we will analyze this question and give two design considerations on what a `good' error pattern is.



\subsection{Consideration Regarding Logical Operation}

\begin{figure}[t]
    \centering
    \includegraphics[width=0.9\linewidth]{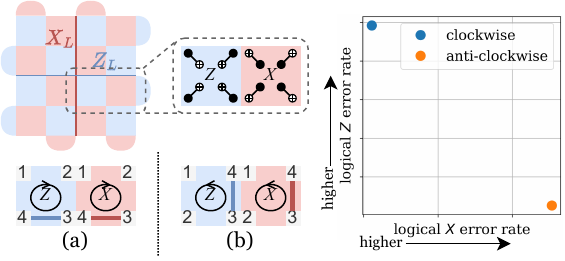}
    \caption{(a) \textbf{clockwise} and (b) \textbf{anti-clockwise} measurement order of a rotated surface code; and their logical $X,Z$ error rates. The numbers represent the order in which the Pauli check is executed between the ancilla qubit in the center and the corresponding data qubit.}
    \Description{(a) \textbf{clockwise} and (b) \textbf{anti-clockwise} measurement order of a rotated surface code; and their logical $X,Z$ error rates}
    \label{fig:twosurfaceorder}
    \vspace{-8pt}
\end{figure}

Our first insight is that the scheduling of the syndrome measurement circuit should let the error propagate to a pattern that is \textit{not close to a logical operation}.
We explain this with the rotated surface code example in Figure~\ref{fig:twosurfaceorder}.
Here, we execute all the Pauli checks of each stabilizer in the same order, and we tested two orders: (a) clockwise and (b) anti-clockwise, starting from the top left data qubit. The orders are similar, but the logical $X$ and $Z$ error rates on the data qubits after syndrome measurement, as shown in Figure~\ref{fig:twosurfaceorder} (decoded by minimal-weight perfect matching~\cite{higgott2025blossom}), are significantly biased. The clockwise order has a low $X$ and high $Z$ logical error rate, and the anti-clockwise order is the opposite.



The reason for this is that in the worst case, either the second half of the data qubits (tick $3,4$) or the first half (tick $1,2$) are affected by hook errors (marked by darker bars in Figure~\ref{fig:twosurfaceorder} (a) and (b)). In the clockwise order, ticks $1-2$ and $3-4$ align with the logical $Z$ operator in $Z$ stabilizer syndrome measurement (horizontally).
Horizontally propagated $Z$ errors are closer to a logical $Z$ operator, leading to a higher logical $Z$ error rate.
Similarly, in the anti-clockwise order, ticks $1-2$ and $3-4$ align with the logical $X$ operator in the $X$ stabilizer syndrome measurement.
The vertically propagated $X$ errors cause a higher logical $X$ error rate.
The example suggests that the order should place late Pauli checks ``orthogonal'' to the corresponding logical operations, which is the insight behind Google's syndrome measurement circuit schedule on surface code.
In Google's schedule (see Figure~\ref{fig:motivation}), the physical $Z$ errors are most likely to propagate vertically (ticks 3 and 4 in the blue Z-type stabilizer), which is perpendicular to the direction of logical $Z$ operations.
The physical $X$ errors are likely to propagate horizontally, which is also perpendicular to the direction of logical $X$ operations.

\textbf{Hardness in considering logical operations:} Generally, for QECC other than the surface code, it is usually hard to have an analytical guideline on how to let the error propagate so that the error pattern is less likely to be close to a logical operation.
For a $[[n,k,d]]$ QECC, there are $k2^{n-k+1}$ different implementations of valid logical $X_L$ or $Z_L$ operations over the $k$ logical qubits.
It is highly non-trivial to determine whether an error pattern is close to one of these logical operations in this exponentially large set.


\subsection{Consideration Regarding Decoder}\label{sec:non-ideal-decoder}

\begin{figure}[t]
    \centering
    \includegraphics[width=0.5\linewidth]{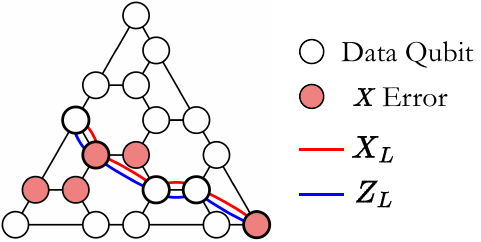}
    \caption{$X$ errors happen on the red data qubits. The error pattern is interpreted differently by the hypergraph unionfind and BP-OSD decoder. Red and blue line marks one logical $X$ and $Z$ operators. Each polygon is an $X$ and $Z$ stabilizer.}
    \Description{..}
    \label{fig:misdecode}
    \vspace{-8pt}
\end{figure}

Our second insight is that the scheduling of the syndrome measurement circuit should let the error propagate to a pattern that \textit{the classical decoder can decode correctly}.
This comes from the fact that practical decoders only give approximate decoding solutions.
It has been proven that generally QECC decoding is an NP-hard problem~\cite{hsieh2011nphardness}, and hardware platforms usually put strict time constraints on how long a decoder can execute to decode one round of syndrome measurement results.
In practice, a classical decoder can only employ a heuristic algorithm to solve this NP-hard decoding problem with an approximate solution.
A heuristic algorithm can not be good at all inputs when solving an NP-hard problem.
It will inevitably provide good solutions for some inputs, but bad solutions for others.
Therefore, a decoder should be able to correctly decode some error patterns while failing on some other error patterns.
An example is shown in Figure~\ref{fig:misdecode}. For a certain error pattern on the $[[19,1,5]]$ hexagonal color code ($X$ error happens on red data qubits), we tested two different decoders: hypergraph unionfind~\cite{berlekamap1978on,lovasz1972factorization} and BP-OSD~\cite{roffe2020bposd}.
 \verb|stim|~\cite{gidney2021stim} simulation shows that BP-OSD gives the correct output while unionfind fails.


\textbf{Hardness in considering the decoder:} It is crucial to consider the decoder so that the propagation of hook errors through syndrome measurement is more likely to be correctly decoded by the decoder, thereby reducing the logical error rate afterward. For a specific error pattern, we can test the heuristic decoding algorithm to determine whether it decodes this error pattern correctly. However, there are an exponentially large number of possible error patterns, making it impractical to check all of them and summarize analytical guidelines for the syndrome measurement circuit scheduling. In general, it is hard to tell what error patterns are good for a specific decoder because we cannot expect an accurate reference solution for an NP-hard problem.

\subsection{Design Objective}

Bringing the above insights together, our goal is: given a commuting‑stabilizer code with its stabilizer set $\{S_i\}$, a physical noise model, and a heuristic decoder, choose a valid schedule (an ordering and allowed parallelism of Pauli checks and ancilla readouts within a round) that minimizes the expected logical error rate. The schedule should shape hook‑error propagation so that (i) the residual data‑qubit errors after a round are far from the support of any logical operator, and (ii) the resulting error patterns fall in the region that the decoder can decode correctly with high probability.



%% file: tex/4-method.tex
\section{Data-Driven Scheduling with MCTS}\label{sec:mcts}

As discussed above, the scheduling problem is both combinatorial and analytically intractable: hook error propagation depends sensitively on order, proximity to logical operators is code-dependent, decoder behavior is heuristic, and the number of valid schedules grows rapidly with circuit size. In practice, this complexity is amplified by device realities—device non‑uniformity—different physical qubits have different noise variances—so a single rule‑based schedule is unlikely to perform well across all scenarios.
Therefore, our \revdelete{compiler}\revadd{synthesizer} \myCompilerNameSpace adopts a data‑driven Monte Carlo Tree Search (MCTS) scheduler\revadd{. It utilizes a list-based circuit representation and applies MCTS to explore efficiently possible syndrome measurement schedulings. Then, it} \revdelete{that} evaluates schedules via noisy simulation rollouts with the decoder in the loop, capturing the two aforementioned design considerations (distance from logical operators and decoder correctability) while adapting to per‑qubit noise variance. \revadd{The evaluation through simulation is then used to guide the search of MCTS.}

\subsection{Circuit Representation}

To represent the syndrome measurement circuit in \myCompilerName, we split the circuit into several \textit{ticks} $1\leq t_i\leq t_{max}$. In each tick, we either perform several Pauli checks or idle. For a Pauli-$\sigma$ ($\sigma = X\text{ or }Z$) measurement, we use the triplet to denote:
    $(data,ancilla,\sigma) \mapsto t_i$,
indicating that it will be performed between the data qubit $data$, the ancilla qubit $ancilla$ during tick $t_i$. For each stabilizer $S_k=\sigma_n\dots\sigma_1$ that will be measured by the ancilla qubit $a_k$, we can collect a list of triplets $(q_j,a_k,\sigma_j)$ for all $\sigma_j\in S_k\neq I$. As all the Pauli checks commute, we will collect all the triplets, regardless of the sequence, and eventually each triplet will be assigned to a tick $t$.

In the same tick, no qubit can be utilized by two Pauli checks. This means that if we have $(d_i,a_i,\sigma_i)$ and $(d_j,a_j,\sigma_j)$ both assigned to the same tick $t$, then we must have the non-conflict condition $d_i\neq d_j$ and $a_i\neq a_j$.

\begin{figure}[tb]
    \centering
    \includegraphics[width=0.9\linewidth]{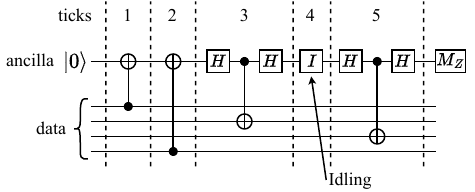}
    \caption{An example syndrome measurement circuit with five ticks. Idlings are not all drawn explicitly.}
    \Description{An example syndrome measurement circuit with five ticks. Idlings are not all drawn explicitly.}
    \label{fig:tickexample}
\end{figure}

As all triplets are assigned to a tick, we can convert it into a concrete quantum circuit by placing the gates at the corresponding tick. If a qubit is not involved in a Pauli check during a tick, then an idling gate $I$ is placed. The example in Figure~\ref{fig:tickexample} shows a five-tick syndrome measurement circuit, with different qubits idling on different ticks. Notice that on the fourth tick, the ancilla qubit is idling.

Once we have the syndrome measurement circuit, we can model the quantum errors, specifically, hook errors, based on hardware information and gate time. At each tick, quantum errors are randomly appended to the ancilla qubits based on the probability (error rate), including idling errors, whose parameters are calculated from $T_1$, $T_2$, and the $CNOT$ time.

To further apply MCTS to schedule syndrome measurement and generate a circuit, we also need to formalize the tree states and an evaluation function for the leaf.

\subsection{Stabilizer Partition}

For stabilizers with $X$ and $Z$ checks on the same data qubit, such as $XZZX$ and $ZXXZ$, their individual Pauli checks cannot be freely swapped due to the anticommutation relationship~\cite{beverland2021cost,geher2024tangling} and they have to be scheduled separately. To address this, we will divide the stabilizers into different partitions using Algorithm~\ref{alg:partition}. 
\begin{algorithm}
    \caption{Partition stabilizers into scheduling groups}
    \label{alg:partition}
    \begin{algorithmic}[1]
        \State $AllPar\gets[]$
        \State $Stab\gets\{\text{all stabilizers}\}$
        \While{$Stab\neq\emptyset$}
            \State $S_0\gets\mathtt{random\_pick}(Stab)$
            \State $Stab.\mathtt{remove}(S_0)$
            \State $Par\gets\{S_0\}$.
            \For{$S\in Stab$}
                \For{$S'\in Par$}
                    \If{$\forall i,\lnot(\sigma_i^S,\sigma_i^{S'}\neq I\land \sigma_i^S\neq \sigma_i^{S'})$}
                        \State $Stab.\mathtt{remove}(S)$
                        \State $Par.\mathtt{add}(S)$
                    \EndIf
                \EndFor
            \EndFor
            \State $AllPar.\mathtt{append}(Par)$
        \EndWhile
        \State $\mathbf{return}\ AllPar$
        \State \textcolor{gray}{// For each $Par\in AllPar$, perform $\mathtt{MCTS}(Par)$}
    \end{algorithmic}
\end{algorithm}

In each partition $Par\in AllPar$, for any two stabilizers $S_i$ and $S_j$, if they have Pauli checks on the same data qubit $q_m$, then they must both perform the same Pauli checks on $q_m$. This means their Pauli checks can be exchanged freely. For example, there are two partitions for CSS codes: all the $X$ stabilizers and all the $Z$ stabilizers. After partition, we perform MCTS on each partition and obtain a list of partial syndrome measurement circuits. The entire schedule is obtained by concatenating these circuits.

\subsection{Scheduling State}

The current state in MCTS represents the state of the scheduling process. We use a (possibly incomplete) list $L$ to record the tick $t$ for each Pauli check $(data,ancilla,\sigma)$. At MCTS leaves, the list is complete, and all Pauli checks are included, with a tick assigned to each. The (possibly incomplete) list should always satisfy the non-conflict condition for Pauli checks within. This is guaranteed when generating possible state transitions.

Starting from a non-final state, we can pick any Pauli check $(q, a, \sigma)$ that is not in the current list $L$ and append it with a minimum possible tick $t$ to get a new state. To obtain $t$, let $t_{max}$ be the max tick of Pauli checks $(q', a',\sigma')$ in the list $L$ that shares the same data or ancilla qubit with $(q, a,\sigma)$. Then, any $t>t_{max}$ ensures the non-conflict condition. However, to minimize the idling error rate and achieve the lowest circuit depth, we pick $t=t_{max}+1$.

Table~\ref{tab:examplelist} shows an example of an incomplete list. When the Pauli check $(q_1, a_2, Z)$ (marked by an arrow) is appended to the list, rows 1 and 3 share the same data or ancilla qubit with it, and their minimum tick is 1. Thus, the new measurement is assigned to tick $2 = 1 + 1$.
\begin{table}[tb]
    \centering
    \caption{Example of MCTS state (incomplete).}
    \label{tab:examplelist}
    \begin{tabular}{c|c|c|c|c|}\cline{2-5}
         & $tick$ & $data$ & $ancilla$ & $\sigma$ \\\cline{2-5}
         & 1 & 1 & 1 & $X$ \\\cline{2-5}
         & 2 & 2 & 1 & $X$ \\\cline{2-5}
         & 1 & 2 & 2 & $Z$ \\\cline{2-5}
         $\mathbf{\rightarrow}$ & 2 & 1 & 2 & $Z$ \\\cline{2-5}
         & $\dots$ &  $\dots$ &  $\dots$ &  $\dots$
    \end{tabular}
\end{table}

\subsection{Evaluation Function}\label{sec:evaluation-function}

To provide the necessary evaluation score for each final state in MCTS, we must build a numerical evaluation function for a complete syndrome measurement circuit, taking into account the decoder.

\begin{figure}[b]
    \centering
    \includegraphics[width=0.8\linewidth]{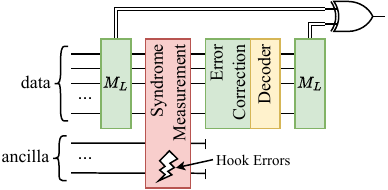}
    \caption{Sampling circuit to calculate logical $X$ (or $Z$) error rate. Green indicates the component is error-free; red indicates that hook errors have been injected.}
    \Description{Sampling circuit to calculate logical $X$ (or $Z$) error rate. Green indicates the component is error-free; red indicates that hook errors have been injected.}
    \label{fig:samplecircuit}
\end{figure}

MCTS takes a higher score if the final state is ``better.'' Based on our optimization goal, we take the inverse of the overall logical error rate ${1}/({1-(1-p_X)(1-p_Z)})$ as the score, where $p_X$ and $p_Z$ are the overall logical $X$ and $Z$ error rates, obtained by numerical simulation over a sampling circuit with \verb|stim|~\cite{gidney2021stim}, as shown in Figure~\ref{fig:samplecircuit}: (1) The circuit first measures all logical $X$ (or $Z$) operators, (2) then performs a syndrome measurement with our custom order and hook errors. (3) After that, an ideal error correction is performed with a designated decoder and an error-free syndrome measurement. (4) We eventually measure all the logical $X$ (or $Z$) values again and compare them with the first round of values, determining if a corresponding logical $Z$ (or $X$) happened by \verb|stim|'s \verb|OBSERVABLE_INCLUDE| instruction.

If a QECC has multiple commuting logical $X$ (or $Z$) operators, we measure all of them in steps (1) and (4). We will sample each circuit multiple times and calculate the error rate based on the number of logical flipping events that occur.

\subsection{Continuous Searching}

A complete MCTS process typically begins with a state, referred to as the \textit{initial state}, and explores different paths of moves to leaves for a specified number of iterations. For each path to a leaf, we calculate the evaluation score of the leaf and propagate backward through the path. Eventually, it will provide the best \textit{next move}. However, our goal is to develop an entire syndrome measurement schedule, which is the final state. If we continuously apply MCTS for each step, then it requires a massive search cost, which takes $\#measurements\times\#iters\_pre\_step$ total explorations.

\begin{figure}[ht]
    \centering
    \includegraphics[width=\linewidth]{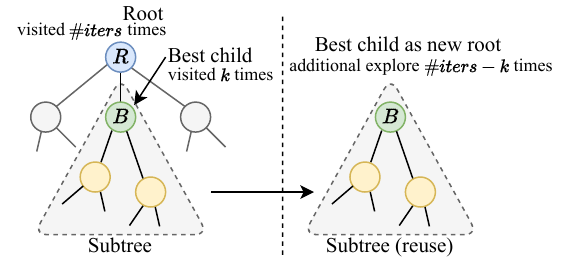}
    \caption{Reuse search subtree in continuous MCTS}
    \Description{Reuse search subtree in continuous MCTS}
    \label{fig:subtree}
\end{figure}

This redundancy can be reduced by recognizing that most search trees share overlapping elements. That is, every time we move forward a step, instead of starting a new search from the next state, we can reuse the subtree from the previous step and continue searching from the subtree root. As shown in Figure~\ref{fig:subtree}, in the last step, we initiated an MCTS from the root $R$ and performed $\#iters$ explorations. Now, the root $R$ has been visited $\#iters$ times and the algorithm selects the best child $B$, which has been visited $k$ times, as the next move. Instead of initiating another search starting from $B$, we take the last subtree with root $B$ (gray triangle) as a new tree and perform another $\#iters-k$ explorations from $B$. In this way, the last exploration paths are reserved and the required iterations are reduced from $\#iters$ to $\#iters-k$.

%% file: tex/5-evaluation.tex
\section{Evaluation}

In this section, we evaluated \myCompilerName{} against multiple baselines on QECCs with different decoders.

\subsection{Experiment Setup}

\subsubsection{QECCs and Decoders}

We select multiple QECCs and decoders as the benchmarks, ranging across codes that encode single or multiple logical qubits. They include: rotated surface code~\cite{bombin2007optimal,horsman2012surface}, color code~\cite{bombin2006topological,rodriguez2024experimental}, 
surface code with defects~\cite{kitaev2003anyons}, bivariate bicycle (BB) code~\cite{bravyi2024highthreshold}, hyperbolic surface code~\cite{breuckmann2016constructions,albuquerque2009topological}, and hyperbolic color code~\cite{delfosse2013tradeoffs,dasilva2018hyperbolic,vuillot2022quantum}.
For the decoder, we include the minimum-weight perfect matching (MWPM)~\cite{higgott2025blossom}, BP-OSD~\cite{roffe2020bposd}, and hypergraph unionfind~\cite{berlekamap1978on,lovasz1972factorization} decoders.

\subsubsection{Error Model}

Unless otherwise specified, we apply an error model adapted from the IBM Brisbane~\cite{ibmquantumplatform} quantum computer in most experiments in this section. $CNOT$ introduces a depolarizing error with $p=0.0074$, and idling at each tick introduces a depolarizing error with $p=0.0052$. The error rates are uniform for all physical qubits.
We use projected lower error rates in error-scaling experiments and introduce variance in the physical error rates for different qubits in the non-uniform error model experiments.

\subsubsection{Machine and Software}

We implemented our algorithm in Python and executed it on a server equipped with two Intel Xeon 6960P processors (72 cores per CPU) and 1.5 TB of RAM. The numerical simulation with \verb|stim| is paralleled with 32 processes. The iteration per MCTS step varies from 4000 to 8000, depending on the code distance.

\input{tex/99-bigtable}

\subsection{Baselines}

We compared \myCompilerName{} against several baseline schedules, including the \textbf{lowest depth schedules}~\cite{kang2025quits,menon2025magic} in general and the known \textbf{industry hand-crafted schedules} for specific codes.

\subsubsection{Low Depth Schedule~\cite{kang2025quits,menon2025magic}}

For arbitrary stabilizer codes, finding their lowest-depth syndrome measurement schedule is a combinatorial optimization problem.
In this paper, we formulate it via integer programming (IP)~\cite{christos1982combinatorial} and solve it via the IP solver in the \verb|pulp| package~\cite{john2024cbc}.

We briefly introduce our problem encoding and formulation. For each Pauli check $(data_i,ancilla_i,\sigma_i)$, we assign an integer variable $t_i>0$ representing its tick in the circuit. and the optimization goal is to minimize $t_{\max}=\max(t_i)$. In order to express $t_{\max}$, we introduce a set of constraints:
    $\forall i,t_{\max}\geq t_i$
where $\geq$ is implemented by the Big M method~\cite{igor2009linear}. 
We also need to apply the non-conflict conditions, that is, no qubit will be used simultaneously in the same tick. Suppose $(data_i,ancilla_i,\sigma_i)$ and $(data_j,ancilla_j,\sigma_j)$ are assigned at tick $t_i$ and $t_j$. If we have $data_i=data_j$ or $ancilla_i=ancilla_j$, then the solution must satisfy $t_i\neq t_j$. The scheduler will iterate over all pairs of Pauli checks and add these constraints.

For the anticommutation relationship~\cite{beverland2021cost,geher2024tangling}, the following contraints are appended if two stabilizers $S_1$ and $S_2$ have anticommute checks on the overlapped data qubits $\{q_m\}$:
\begin{equation*}
    \prod_{\forall q_i,\in\{q_m\}}(t^{S_1}_i-t^{S_2}_i)>0
\end{equation*}

The solver will find a schedule with the guaranteed lowest depth or an approximation before the timeout (\textbf{1 day}).

\subsubsection{Industry Hand-Crafted Schedules}
We also evaluate the industry hand-crafted schedule for the rotated surface code and the BB code, as they are actively pursued by Google and IBM, respectively.
\textbf{Google}'s zig-zag schedule~\cite{acharya2024qecsurface,horsman2012surface} for the rotated surface code can be found at Figure~\ref{fig:motivation} in Section~\ref{sec:introduction}.
\textbf{IBM}'s schedule can be found in~\cite{bravyi2024highthreshold,bravyi2024bivariate,yoder2025tour}.

\subsection{Comparing with Lowest Depth Schedules}

\subsubsection{\revadd{Logical Error Rate Reduction}}

As shown in Table~\ref{tab:overall-error}, \myCompilerName{} reduces the \emph{overall} logical error rate by \textbf{80.6\%} on average across all 32 code/decoder instances of different code families, code sizes, and decoding algorithms, with a peak reduction of \textbf{96.2\%}. Although \myCompilerName{} typically generates syndrome‑measurement schedules with greater depth than the lowest‑depth baseline, the net reliability still improves substantially because the learned ordering \emph{tunes hook‑error propagation}, steering faults into patterns that are away from logical operations and more readily corrected by the decoder.
Note that $[[9,1,3]]$ XZZX code is the only case where \myCompilerName{} shows similar performance with the baseline. This is because the code is small, allowing the baseline to hit a good schedule in its small design space.

\subsubsection{\revadd{Space-Time Resource Estimation and Reduction}}\label{sec:runtime-resource}

\revadd{
Although \myCompilerName{} may produce syndrome-measurement schedules with moderately increased circuit depth, its substantially lower logical error rates reduce the required code distance to achieve a target reliability. Since the execution cost of fault-tolerant quantum computation is dominated by repeated syndrome-measurement cycles~\cite{fowler2012surface,litinski2019game,gidney2021factor}, we evaluate system-level impact using the space--time volume, defined as the product of physical qubit count and execution time per syndrome-measurement round.
}

\revadd{
To make a fair comparison, we normalize all configurations by the achieved logical error rate rather than by fixed code parameters. Specifically, based on the logical error rates reported in Table~\ref{tab:overall-error}, we select, from the hexagonal color code, square-octagonal color code, and hyperbolic surface code families, one configuration produced by \myCompilerName{} and one lowest-depth baseline configuration that achieves the same or very similar logical error rates. Because baseline schedules typically yield higher error rates per round, they must employ larger code distances—and thus more physical qubits—to reach comparable reliability.
}

\begin{table}[t]
    \centering
    \caption{\revadd{Space-time volume comparison at comparable logical error rates}}
    \label{tab:resource-estimate}
    \resizebox{0.95\linewidth}{!}{
    \begin{tabular}{|c|c|c|c|c|c|}\hline
          & \Centerstack{$[[n,k,d]]$ \\ \& Depth} & \Centerstack{Logical \\ Error Rate} & \Centerstack{Time \\ $/ms$} & \Centerstack{Space-time \\ Volume \\ $/ms\times\#q$} \\\hline
         \multicolumn{5}{|c|}{\textbf{Hexagonal Color Code, BP-OSD}} \\\hline
         \myCompilerName{} & [[7,1,3]], 14 & $1.04\times10^{-3}$ & 12.4 & 86.8\\\hline
         Lowest Depth & [[61,1,9]], 15 & $2.23\times10^{-3}$ & 13.0 & 793.0 \\\hline
         \multicolumn{4}{|c|}{\textbf{Reduction}} & 89.0\% \\\hline\hline
         
         \multicolumn{5}{|c|}{\textbf{Square-Octagonal Color Code, BP-OSD}} \\\hline
         \myCompilerName{} & [[7,1,3]], 13 & $9.5\times10^{-4}$ & 11.8 & 82.6\\\hline
         Lowest Depth & [[49,1,9]], 15 & $2.39\times10^{-3}$ & 13.0 & 637.0 \\\hline
         \multicolumn{4}{|c|}{\textbf{Reduction}} & 87.0\% \\\hline\hline

         \multicolumn{5}{|c|}{\textbf{Hyperbolic Surface Code, MWPM}} \\\hline
         \myCompilerName{} & [[36,8,4]], 16 & $1.31\times10^{-2}$ & 13.6 & 489.6\\\hline
         Lowest Depth & [[60,8,4]], 10 & $1.16\times10^{-2}$ & 10.0 & 600.0 \\\hline
         \multicolumn{4}{|c|}{\textbf{Reduction}} & 18.4\% \\\hline
    \end{tabular}
    }
\end{table}

We estimate absolute execution time using the IBM Brisbane device model, where a two-qubit gate costs approximately $600$ ns and a measurement takes approximately $4000$ ns~\cite{zheng2024minimizing}. Single-qubit gate latency is negligible at this scale. The execution time of one syndrome-measurement round is therefore computed as:
    $T_{\text{round}} = \#\text{depth} \times T_{2Q} + T_{\text{meas}}$,
and the corresponding space-time volume is given by 
    $\text{Space-Time Volume} = T_{\text{round}} \times \#\text{physical qubits}$.
Table~\ref{tab:resource-estimate} reports both the absolute execution time and the resulting space--time volume. While \myCompilerName{} sometimes incurs a higher depth in terms of two-qubit gates, this does not significantly increase—and in some cases even reduces—the absolute execution time required to achieve a given logical error rate. This effect arises from two factors. First, to reach comparable reliability, lowest-depth baseline schedules require larger code distances, whose minimal-depth schedules themselves have larger circuit depths. Second, syndrome-measurement rounds are dominated by the ancilla measurement latency, which is substantially longer than two-qubit gate latency (this holds for most hardware platforms); as a result, moderate increases in two-qubit gate depth are largely amortized by the fixed measurement cost.

\revadd{
Overall, \myCompilerName{} reduces the required space--time volume by 20-90\% across the evaluated code families. The dominant contributor to this reduction is the ability to operate at smaller code distances, which significantly lowers the physical qubit count, while absolute execution time is comparable to or smaller than that of lowest-depth baselines at similar logical error rates. These results confirm that optimizing syndrome-measurement scheduling for logical reliability can yield substantial system-level benefits without increasing runtime requirements.
}

\begin{figure}[t]
    \centering
    \begin{minipage}{\linewidth}
        \centering
        \includegraphics[width=0.93\linewidth]{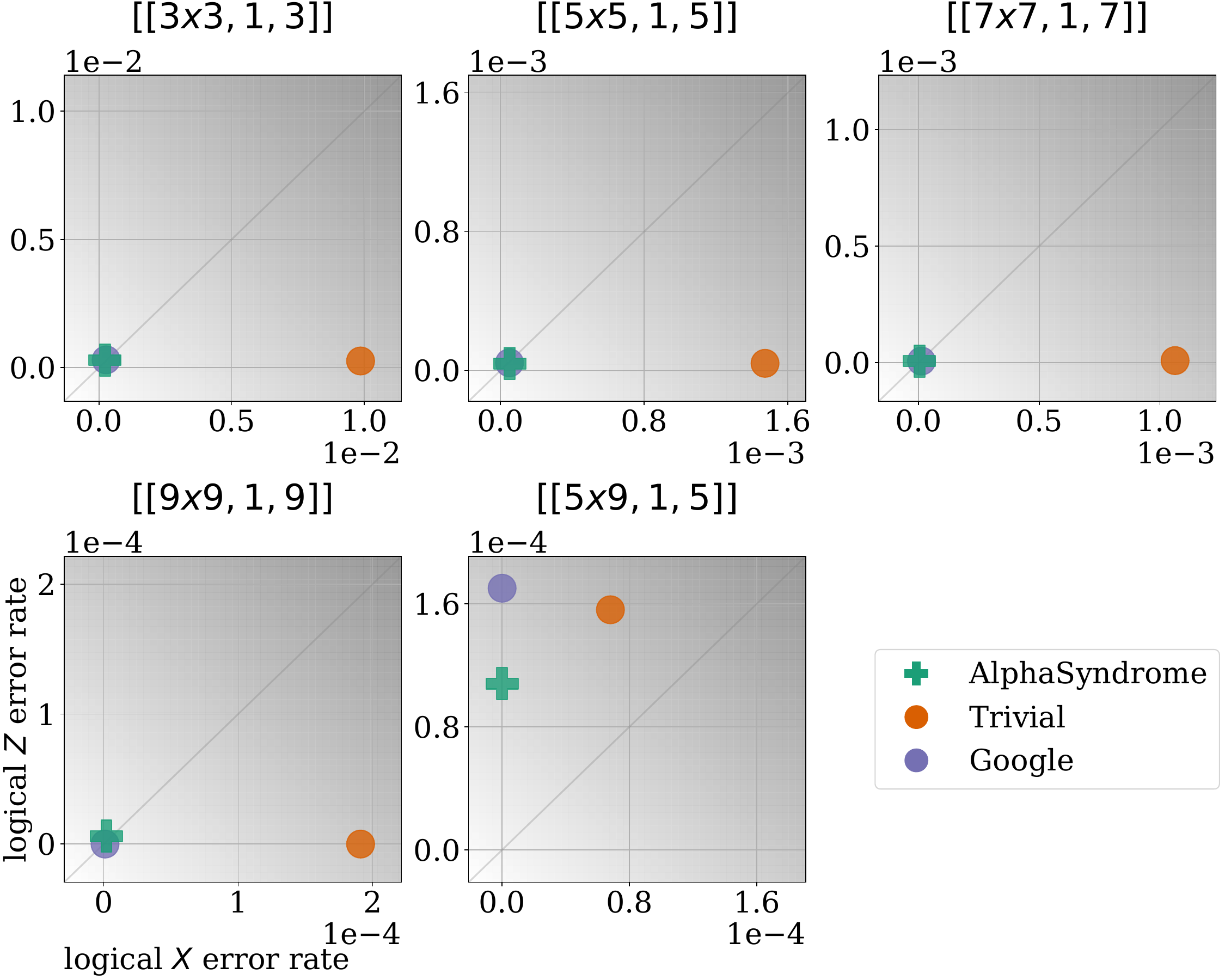}
        \caption{Logical $X$ and $Z$ error rates of \myCompilerName{} against \textbf{Google}'s schedule on rotated surface codes.}
        \Description{..}
        \label{fig:surface-result}
    \end{minipage}
    \begin{minipage}{\linewidth}
        \centering
        \vspace{5pt}
        \includegraphics[width=0.93\linewidth]{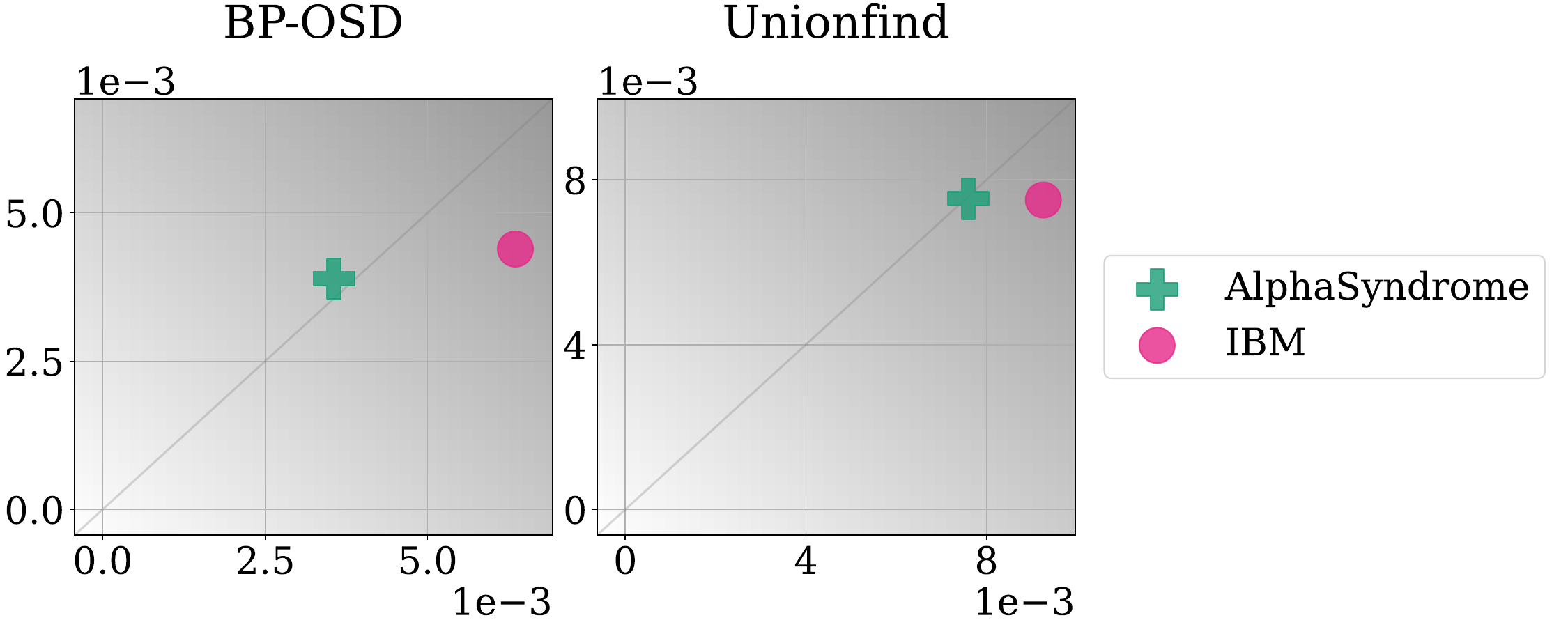}
        \caption{Logical $X$ and $Z$ error rates of \myCompilerName{} against \textbf{IBM}'s schedule on $[[72,12,6]]$ BB code.}
        \Description{Logical $X$, $Z$ and overall error rate of $[[72,12,6]]$ bivariate bicycle code test case.}
        \label{fig:bbcode}
    \end{minipage}
    \vspace{-10pt}
\end{figure}

\begin{table*}[t]
    \centering
    \caption{Overall logical error rate of cross-testing schedules compiled under the BP-OSD and hypergraph unionfind decoders.}
    \label{tab:cross-decoder}
     \resizebox{0.75\textwidth}{!}{
    \begin{tabular}{|c|c|c|c||c|c|c|}\hline
\textbf{Testing decoder} & \multicolumn{3}{c||}{\textbf{BP-OSD}} & \multicolumn{3}{c|}{\textbf{Unionfind}} \\\hline
\textbf{Scheduling decoder}& \textbf{BP-OSD} & Unionfind & $\leftarrow$ Reduction & BP-OSD & \textbf{Unionfind} & $\rightarrow$ Reduction \\\hline

\hline\multicolumn{7}{|c|}{\textbf{Hexagonal Color Code}}\\\hline
$[[7,1,3]]$ & $\mathbf{1.03\times 10^{-3}}$ & $1.36\times 10^{-2}$ & $92.43\%$ & $9.56\times 10^{-4}$ & $\mathbf{8.22\times 10^{-4}}$ & $14.01\%$\\\hline
$[[19,1,5]]$ & $\mathbf{1.1\times 10^{-3}}$ & $1.32\times 10^{-3}$ & $16.54\%$ & $3.01\times 10^{-3}$ & $\mathbf{1.48\times 10^{-3}}$ & $50.78\%$\\\hline
$[[37,1,7]]$ & $\mathbf{1.88\times 10^{-4}}$ & $3.9\times 10^{-4}$ & $51.79\%$ & $1.06\times 10^{-3}$ & $\mathbf{7.62\times 10^{-4}}$ & $28.24\%$\\\hline
$[[61,1,9]]$ & $1.32\times 10^{-4}$ & $\mathbf{1.12\times 10^{-4}}$ & $-17.86\%$ & $5.24\times 10^{-4}$ & $\mathbf{3.0\times 10^{-4}}$ & $42.75\%$\\\hline
\hline\multicolumn{7}{|c|}{\textbf{Square-Octagonal Color Code}}\\\hline
$[[7,1,3]]$ & $9.58\times 10^{-4}$ & $\mathbf{8.5\times 10^{-4}}$ & $-12.70\%$ & $9.18\times 10^{-4}$ & $\mathbf{8.14\times 10^{-4}}$ & $11.33\%$\\\hline
$[[17,1,5]]$ & $\mathbf{1.18\times 10^{-3}}$ & $1.61\times 10^{-3}$ & $26.77\%$ & $2.57\times 10^{-3}$ & $\mathbf{1.93\times 10^{-3}}$ & $25.10\%$\\\hline
$[[31,1,7]]$ & $\mathbf{3.96\times 10^{-4}}$ & $5.78\times 10^{-4}$ & $31.48\%$ & $1.88\times 10^{-3}$ & $\mathbf{1.0\times 10^{-3}}$ & $46.58\%$\\\hline
$[[49,1,9]]$ & $\mathbf{1.86\times 10^{-4}}$ & $2.18\times 10^{-4}$ & $14.68\%$ & $1.19\times 10^{-3}$ & $\mathbf{5.3\times 10^{-4}}$ & $55.53\%$\\\hline

    \end{tabular}
     }
\end{table*}

\subsection{Comparing with Industry Hand-Crafted Schedules}

Figure~\ref{fig:surface-result} shows the comparison among \myCompilerName{}, \textbf{Google}'s schedule, and the \textbf{lowest depth} baseline for the rotated surface code with MWPM decoder.
It can be observed that \myCompilerName{}'s automatically generated schedule can outperform the \textbf{lowest depth} schedule and match the performance of \textbf{Google}'s schedule across different code sizes and distances, even for rectangular rotated surface code with biased error correction capability.

Figure~\ref{fig:bbcode} shows the comparison between \myCompilerName{} and \textbf{IBM}'s schedule for the $[[72,12,6]]$ BB code with both BP-OSD and Unionfind decoder.
\myCompilerName{} achieved a 44\% reduction with BP-PSD, and 10\% reduction with Unionfind in overall logical error rate compared to \textbf{IBM}.


\subsection{Cross-Decoder Evaluation}

In this experiment, we want to test and confirm whether \myCompilerName{} can \emph{tailor} its schedule to different decoders: our hypothesis is that a schedule searched with a decoder performs best when evaluated with the same decoder and typically degrades when evaluated with a different decoder. 
The cross‑decoder study below is designed to validate this specialization.
We compile eight instances of color code using both BP-OSD and Unionfind decoders, and then cross-test on the other decoder for the logical error rate.

Table~\ref{tab:cross-decoder} shows that \myCompilerName{} tailors its schedule to the target decoder. When tested with BP‑OSD, schedules compiled using BP‑OSD reduce the overall logical error rate by \textbf{25.4\%} on average compared to schedules compiled with unionfind, winning in \textbf{7/8} instances and peaking at \textbf{92.4\%} reduction on $[[7,1,3]]$. Symmetrically, when tested with \emph{unionfind}, schedules compiled using unionfind reduce errors by \textbf{34.3\%} on average over BP‑OSD‑compiled schedules, again in \textbf{8/8} instances, with a peak of \textbf{55.53\%}. Overall, cross‑testing confirms that using a schedule compiled for a different decoder generally degrades performance, indicating that \myCompilerName{} indeed optimizes towards decoder behavior.

\subsection{Low Physical Error Rate Evaluation}

\begin{figure*}[h]
    \centering
    \includegraphics[width=0.97\linewidth]{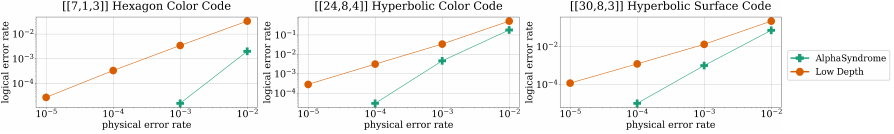}
    \caption{Performance of \myCompilerName{} under low physical error rate.}
    \label{fig:scaling}
    \Description{..}
\end{figure*}

In this experiment, we scale down the physical error rate to test whether \myCompilerName{} is still effective under low physical error settings.
We set the $CNOT$ and idling error rates to $10^{-2}$, $10^{-3}$, $10^{-4}$, and $10^{-5}$, and \revdelete{compile}\revadd{synthesize} the syndrome measurement circuits of three QECCs.
As shown in Figure~\ref{fig:scaling}, \myCompilerName{} is capable to learn the error pattern even when the physical error rate is down to $10^{-5}$, and consistently achieves a better result than the \textbf{lowest depth} baseline.
In particular, the trend in Figure~\ref{fig:scaling} shows that our logical error rate reduction is even more significant as the physical error rate decreases.

\subsection{Non-Uniform Error Model}

\begin{figure}[t]
    \centering
    \includegraphics[width=0.95\linewidth]{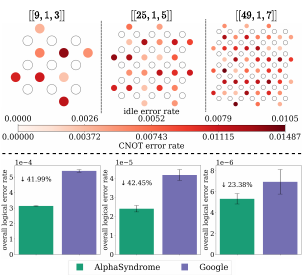}
    \caption{\myCompilerName{} versus \textbf{Google}'s schedule under a non-uniform error model. Upper half: non-uniform error rates on ancilla qubits. Lower half: Overall logical error rates.}
    \Description{..}
    \label{fig:non-uniform}
\end{figure}

In the last experiments, we evaluate \myCompilerName{} with non-uniform error models. We do not compare with the lowest depth baseline since \myCompilerName{} has already outperformed it under the uniform error models.
We only compared with \textbf{Google}'s schedule on the rotated surface code (MWPM decoder), where \myCompilerName{} demonstrated matching performance.
We add some variance to the error rates in IBM Brisbane's base model, as shown in the upper half of Figure~\ref{fig:non-uniform}. 
The color shows the related error rates of the ancilla qubits, on the $[[9,1,3]]$, $[[25,1,5]]$, and $[[49,1,7]]$ rotated surface codes.

As shown on the lower half of Figure~\ref{fig:non-uniform},  \myCompilerName{} can outperform \textbf{Google}'s schedule with significant reduction in logical error rates 
The reason is that \textbf{Google}'s schedule on the rotated surface codes is designed based on a uniform error model, assuming all qubits are identical regarding error probability. Its performance degrades in a non-uniform error model because it does not account for the uneven weights in the MWPM decoder resulting from the non-uniform error model.
Instead, \myCompilerName{}'s data-driven \revdelete{compilation}\revadd{synthesis} can automatically tailor the scheduling to different error models and decoders with different weights.

%% file: tex/99-bigtable.tex
\begin{table*}[t]
\centering
\caption{Logical error rates and generated circuit depth of \myCompilerName{} against the low depth schedule.
$Err_X$, $Err_Z$ and \textbf{Overall} represent the logical $X$, $Z$ and overall error rates, respectively. For the QECCs with multiple logical qubits in one code block, these rates count for cases where at least one logical qubit has an error. Bold marks the lower error rate.}
\label{tab:overall-error}
\resizebox{0.94\textwidth}{!}{
\begin{tabular}{|c|c|c|c|c|c|c|c|c|c|c|}
\hline
\multirow{2}{*}{$\mathbf{[[n,k,d]]}$} & \multirow{2}{*}{\textbf{Decoder}} & \multicolumn{4}{c|}{\textbf{\myCompilerName}}                                                         & \multicolumn{4}{c|}{\textbf{Low Depth}} & \multirow{2}{*}{\shortstack{\textbf{Overall} \\ \textbf{Reduction}}}                                                            \\ \cline{3-10} 
                                      &                                   & \multicolumn{1}{c|}{\textbf{$\mathbf{Err_X}$}} & \multicolumn{1}{c|}{\textbf{$\mathbf{Err_Z}$}} & \textbf{Overall} & \textbf{Depth} & \multicolumn{1}{c|}{\textbf{$\mathbf{Err_X}$}} & \multicolumn{1}{c|}{\textbf{$\mathbf{Err_Z}$}} & \textbf{Overall} & \textbf{Depth} & \\ \hline\hline

\multicolumn{11}{|c|}{\textbf{Hexagonal Color Code}} \\\hline
\multirow{2}{*}{$[[7,1,3]]$} & BP-OSD & $\mathbf{4.64\times 10^{-4}}$ & $\mathbf{5.78\times 10^{-4}}$ & $\mathbf{1.04\times 10^{-3}}$ & 14 & $5.3\times 10^{-3}$ & $1.67\times 10^{-2}$ & $2.2\times 10^{-2}$ & 7 & 95.26\% \\\cline{2-11}
 & Unionfind & $\mathbf{4.12\times 10^{-4}}$ & $\mathbf{4.56\times 10^{-4}}$ & $\mathbf{8.68\times 10^{-4}}$ & 11 & $5.23\times 10^{-3}$ & $1.67\times 10^{-2}$ & $2.18\times 10^{-2}$ & 7 & 96.02\% \\\hline
\multirow{2}{*}{$[[19,1,5]]$} & BP-OSD & $\mathbf{4.49\times 10^{-4}}$ & $\mathbf{4.85\times 10^{-4}}$ & $\mathbf{9.34\times 10^{-4}}$ & 14 & $2.13\times 10^{-3}$ & $3.01\times 10^{-3}$ & $5.14\times 10^{-3}$ & 7 & 81.82\% \\\cline{2-11}
 & Unionfind & $\mathbf{7.53\times 10^{-4}}$ & $\mathbf{6.88\times 10^{-4}}$ & $\mathbf{1.44\times 10^{-3}}$ & 17 & $2.38\times 10^{-3}$ & $3.26\times 10^{-3}$ & $5.63\times 10^{-3}$ & 7 & 74.42\% \\\hline
\multirow{2}{*}{$[[37,1,7]]$} & BP-OSD & $\mathbf{1.24\times 10^{-4}}$ & $\mathbf{1.32\times 10^{-4}}$ & $\mathbf{2.56\times 10^{-4}}$ & 18 & $1.48\times 10^{-3}$ & $2.38\times 10^{-3}$ & $3.86\times 10^{-3}$ & 13 & 93.37\% \\\cline{2-11}
 & Unionfind & $\mathbf{4.26\times 10^{-4}}$ & $\mathbf{3.36\times 10^{-4}}$ & $\mathbf{7.62\times 10^{-4}}$ & 22 & $2.7\times 10^{-3}$ & $3.28\times 10^{-3}$ & $5.97\times 10^{-3}$ & 13 & 87.24\% \\\hline
\multirow{2}{*}{$[[61,1,9]]$} & BP-OSD & $\mathbf{5.6\times 10^{-5}}$ & $\mathbf{6.3\times 10^{-5}}$ & $\mathbf{1.19\times 10^{-4}}$ & 21 & $9.51\times 10^{-4}$ & $1.28\times 10^{-3}$ & $2.23\times 10^{-3}$ & 15 & 94.66\% \\\cline{2-11}
 & Unionfind & $\mathbf{1.89\times 10^{-4}}$ & $\mathbf{1.13\times 10^{-4}}$ & $\mathbf{3.02\times 10^{-4}}$ & 19 & $2.83\times 10^{-3}$ & $3.31\times 10^{-3}$ & $6.13\times 10^{-3}$ & 15 & 95.08\% \\\hline

\multicolumn{11}{|c|}{\textbf{Square-Octagonal Color Code}} \\\hline
\multirow{2}{*}{$[[7,1,3]]$} & BP-OSD & $\mathbf{4.35\times 10^{-4}}$ & $\mathbf{5.15\times 10^{-4}}$ & $\mathbf{9.5\times 10^{-4}}$ & 13 & $5.46\times 10^{-3}$ & $1.68\times 10^{-2}$ & $2.22\times 10^{-2}$ & 7 & 95.72\% \\\cline{2-11}
 & Unionfind & $\mathbf{3.83\times 10^{-4}}$ & $\mathbf{4.07\times 10^{-4}}$ & $\mathbf{7.9\times 10^{-4}}$ & 12 & $5.46\times 10^{-3}$ & $1.36\times 10^{-2}$ & $1.9\times 10^{-2}$ & 7 & 95.85\% \\\hline
\multirow{2}{*}{$[[17,1,5]]$} & BP-OSD & $\mathbf{5.5\times 10^{-4}}$ & $\mathbf{5.6\times 10^{-4}}$ & $\mathbf{1.11\times 10^{-3}}$ & 17 & $6.7\times 10^{-3}$ & $2.57\times 10^{-3}$ & $9.25\times 10^{-3}$ & 9 & 88.00\% \\\cline{2-11}
 & Unionfind & $\mathbf{9.46\times 10^{-4}}$ & $\mathbf{9.52\times 10^{-4}}$ & $\mathbf{1.9\times 10^{-3}}$ & 18 & $6.11\times 10^{-3}$ & $2.9\times 10^{-3}$ & $8.99\times 10^{-3}$ & 9 & 78.89\% \\\hline
\multirow{2}{*}{$[[31,1,7]]$} & BP-OSD & $\mathbf{2.33\times 10^{-4}}$ & $\mathbf{1.63\times 10^{-4}}$ & $\mathbf{3.96\times 10^{-4}}$ & 19 & $2.97\times 10^{-3}$ & $1.3\times 10^{-3}$ & $4.26\times 10^{-3}$ & 12 & 90.71\% \\\cline{2-11}
 & Unionfind & $\mathbf{4.79\times 10^{-4}}$ & $\mathbf{5.26\times 10^{-4}}$ & $\mathbf{1.0\times 10^{-3}}$ & 20 & $5.93\times 10^{-3}$ & $2.79\times 10^{-3}$ & $8.7\times 10^{-3}$ & 12 & 88.46\% \\\hline
\multirow{2}{*}{$[[49,1,9]]$} & BP-OSD & $\mathbf{1.26\times 10^{-4}}$ & $\mathbf{7.9\times 10^{-5}}$ & $\mathbf{2.05\times 10^{-4}}$ & 19 & $1.31\times 10^{-3}$ & $1.09\times 10^{-3}$ & $2.39\times 10^{-3}$ & 15 & 91.44\% \\\cline{2-11}
 & Unionfind & $\mathbf{2.28\times 10^{-4}}$ & $\mathbf{2.88\times 10^{-4}}$ & $\mathbf{5.16\times 10^{-4}}$ & 20 & $4.83\times 10^{-3}$ & $3.42\times 10^{-3}$ & $8.24\times 10^{-3}$ & 15 & 93.74\% \\\hline

\multicolumn{11}{|c|}{\textbf{Hyperbolic Color Code}} \\\hline
\multirow{1}{*}{$[[24,8,4]]$} & Unionfind & $\mathbf{2.42\times 10^{-2}}$ & $\mathbf{2.43\times 10^{-2}}$ & $\mathbf{4.79\times 10^{-2}}$ & 19 & $6.32\times 10^{-2}$ & $5.07\times 10^{-2}$ & $1.11\times 10^{-1}$ & 11 & 56.73\% \\\hline
\multirow{1}{*}{$[[32,12,4]]$} & Unionfind & $\mathbf{3.56\times 10^{-2}}$ & $\mathbf{3.82\times 10^{-2}}$ & $\mathbf{7.25\times 10^{-2}}$ & 19 & $5.82\times 10^{-2}$ & $7.03\times 10^{-2}$ & $1.24\times 10^{-1}$ & 14 & 41.76\% \\\hline
\multirow{1}{*}{$[[40,16,4]]$} & Unionfind & $\mathbf{6.63\times 10^{-2}}$ & $\mathbf{6.52\times 10^{-2}}$ & $\mathbf{1.27\times 10^{-1}}$ & 25 & $2.33\times 10^{-1}$ & $2.08\times 10^{-1}$ & $3.93\times 10^{-1}$ & 28 & 67.63\% \\\hline

\multicolumn{11}{|c|}{\textbf{Hyperbolic Surface Code}} \\\hline
\multirow{1}{*}{$[[30,8,3]]$} & MWPM & $\mathbf{8.24\times 10^{-3}}$ & $\mathbf{9.26\times 10^{-3}}$ & $\mathbf{1.74\times 10^{-2}}$ & 14 & $2.35\times 10^{-2}$ & $2.01\times 10^{-2}$ & $4.32\times 10^{-2}$ & 8 & 59.65\% \\\hline
\multirow{1}{*}{$[[36,8,4]]$} & MWPM & $\mathbf{3.26\times 10^{-3}}$ & $\mathbf{9.89\times 10^{-3}}$ & $\mathbf{1.31\times 10^{-2}}$ & 16 & $1.51\times 10^{-2}$ & $4.2\times 10^{-2}$ & $5.65\times 10^{-2}$ & 8 & 76.76\% \\\hline
\multirow{1}{*}{$[[40,10,4]]$} & MWPM & $\mathbf{8.26\times 10^{-3}}$ & $\mathbf{8.02\times 10^{-3}}$ & $\mathbf{1.62\times 10^{-2}}$ & 15 & $3.07\times 10^{-2}$ & $3.28\times 10^{-2}$ & $6.25\times 10^{-2}$ & 7 & 74.04\% \\\hline
\multirow{1}{*}{$[[60,18,3]]$} & MWPM & $\mathbf{2.3\times 10^{-2}}$ & $\mathbf{2.13\times 10^{-2}}$ & $\mathbf{4.39\times 10^{-2}}$ & 17 & $5.67\times 10^{-2}$ & $7.57\times 10^{-2}$ & $1.28\times 10^{-1}$ & 13 & 65.76\% \\\hline
\multirow{1}{*}{$[[60,8,4]]$} & MWPM & $\mathbf{6.81\times 10^{-4}}$ & $\mathbf{1.96\times 10^{-3}}$ & $\mathbf{2.64\times 10^{-3}}$ & 17 & $3.38\times 10^{-3}$ & $8.29\times 10^{-3}$ & $1.16\times 10^{-2}$ & 10 & 77.35\% \\\hline
\multirow{1}{*}{$[[80,18,5]]$} & MWPM & $\mathbf{6.8\times 10^{-3}}$ & $\mathbf{7.05\times 10^{-3}}$ & $\mathbf{1.38\times 10^{-2}}$ & 18 & $1.9\times 10^{-2}$ & $2.41\times 10^{-2}$ & $4.26\times 10^{-2}$ & 11 & 67.61\% \\\hline

\multicolumn{11}{|c|}{\textbf{Defect Surface Code}} \\\hline
\multirow{1}{*}{$[[25,2,5]]$} & MWPM & $\mathbf{1.99\times 10^{-3}}$ & $\mathbf{6.0\times 10^{-6}}$ & $\mathbf{1.99\times 10^{-3}}$ & 12 & $5.26\times 10^{-2}$ & $7.7\times 10^{-5}$ & $5.27\times 10^{-2}$ & 5 & 96.22\% \\\hline
\multirow{1}{*}{$[[41,2,7]]$} & MWPM & $\mathbf{1.55\times 10^{-3}}$ & $\mathbf{1.0\times 10^{-6}}$ & $\mathbf{1.55\times 10^{-3}}$ & 15 & $2.12\times 10^{-2}$ & $2.2\times 10^{-5}$ & $2.12\times 10^{-2}$ & 5 & 92.67\% \\\hline

\end{tabular}}
\end{table*}

%% file: tex/6-conclusion.tex
\section{Related Work and Discussion}\label{sec:related-work}

\revadd{
\textbf{Syndrome Measurement Scheduling}}
\revdelete{\textbf{Syndrome Measurement Scheduling}}The syndrome measurement circuit scheduling problem for QEC codes plays an essential role in realizing fault-tolerant quantum computation. Some prior works on QEC discuss circuit scheduling for different QEC codes. For example,  ~\cite{horsman2012surface, tomita2014low-distance, dennis2002topological} examined circuit scheduling for surface code. In ~\cite{tomita2014low-distance}, the authors propose executing the gate following an S- or Z-shape ordering to minimize the effect of hook error. However, beyond surface code, this problem has not been well optimized. For instance, \cite{lacroix2025scaling} adopts a trivial ordering using the qubit index for measuring syndromes in their color code experiments. Similarly, ~\cite{pecorari2025high} also adopts the trivial ordering for their hypergraph product code experiment, though this is a result of an analysis of hook error on effective code distance on hypergraph product code~\cite{manes2025distancepreserving}.
Furthermore, ~\cite{bravyi2024high} designed the syndrome extraction circuit to have minimal circuit depth for the BB code. However, these prior works only focus on a specific family of code, and their manual design approach cannot be automatically extended to other codes. Moreover, some works ~\cite{lacroix2025scaling, pecorari2025high} ignore the influence of hook error and focus only on the overall depth of the circuit. Even for works that consider hook error ~\cite{tomita2014low-distance,manes2025distancepreserving}, their analysis remains at the circuit level without evaluating the logical error rate. Moreover, the approximation in the decoder and the non-uniform errors are not yet considered, leading to performance degradation. 
In contrast, the \myCompilerName{} proposed in this paper can automatically take all these factors into consideration and yield a syndrome measurement circuit implementation with much lower logical error rates.

\revadd{
\textbf{QEC Mapping and Gadget Synthesis}
Several prior works develop synthesis techniques for other components of the QEC stack. QECC-Synth~\cite{yin2025qeccsynth} employs MaxSAT to synthesize QEC layouts on sparse hardware architectures by optimizing qubit placement, routing structures, and ancilla-bridge usage, while Chiew \emph{et al.}~\cite{chiew2024fault} study SWAP-based embedding techniques that map fault-tolerant circuits onto constrained-connectivity hardware while preserving fault-tolerance guarantees. These approaches operate at the mapping and layout layer of the stack and do not modify the internal execution order of stabilizer measurements.
}

\revadd{
SAT-based techniques are also used for QEC \emph{state preparation}. Peham \emph{et al.}~\cite{peham2025automated} apply SAT solving to synthesize fault-tolerant state-preparation and verification circuits for CSS codes. These circuits perform error detection on a limited subset of stabilizers and rely on repeat-until-success execution, which is appropriate for initialization but not for the repeated, full syndrome-measurement cycles required during ongoing fault-tolerant computation.
}

\revadd{
While SAT is a natural choice for scheduling problems with large combinatorial spaces, it is less well-suited to the syndrome-measurement scheduling problem considered here for two reasons. First, a full syndrome-measurement round may contain a large number of commuting Pauli checks, yielding an enormous number of valid schedules; encoding all ordering choices and constraints at this scale quickly leads to poor solver scalability. Second, the performance objective in this setting is not a structural metric such as depth or gate count, but a decoder-conditioned \emph{statistical logical error rate} under realistic noise. This objective depends on stochastic noise propagation and heuristic decoder behavior, and is therefore difficult to express as a tractable symbolic cost function for SAT or MaxSAT optimization.
}

\revadd{
\textbf{MCTS-Based Quantum Synthesis and Routing}
MCTS has been applied in several quantum synthesis and optimization contexts, including AlphaRouter~\cite{wei2024alpharouter} and other MCTS- or reinforcement-learning–based approaches for routing and unitary synthesis~\cite{valcarce2025unitary,rietsch2024unitary}. These methods primarily target NISQ-era circuits and optimize structural metrics such as gate count, circuit depth, or routing overhead. In such settings, \textit{partial constructions admit meaningful intermediate cost estimates} (e.g., remaining routing distance or local gate overhead), enabling effective guidance of the search before a circuit is fully specified.
}

\revadd{
Syndrome-measurement scheduling in fault-tolerant QEC differs in a fundamental way. While the schedule is constructed incrementally, \textit{its performance cannot be reliably assessed until the entire syndrome-measurement round is completed.} The relevant objective—decoder-conditioned statistical logical error rate under realistic noise—emerges only from the global interaction of all scheduled Pauli checks, fault propagation, and decoder behavior, and does not admit accurate partial or local cost estimates. As a result, techniques that rely on informative intermediate metrics are poorly aligned with this problem. AlphaSyndrome therefore applies MCTS in a simulation-guided manner over complete schedules, using full rollouts to evaluate terminal states while restricting the search space to commutation-preserving reorderings that maintain fault-tolerance guarantees. 
}

%% file: tex/7-acknowledgement.tex
\section*{Acknowledgements}

We thank the anonymous reviewers for their constructive feedback.
This work was supported in part by the U.S. Department of Energy, Office of Science, Office of Advanced Scientific Computing Research under Contract No. DE-AC05-00OR22725 through the Accelerated Research in Quantum Computing Program MACH-Q project, the U.S. National Science Foundation CAREER Award No. CCF-2338773, and ExpandQISE Award No. OSI 2427020. 

%% file: tex/99-appendix.tex
\appendix

\section{Artifact Appendix}

\newcommand{\doiurl}{\url{https://doi.org/10.5281/zenodo.18291927}}

\subsection{Abstract}

This artifact contains the scheduling algorithm of \myCompilerName{}, tested QECCs, scheduling results, and evaluation scripts for reproducing key results in the paper, including all tables and figures. The software is implemented in Python and provides instructions to prepare, evaluate, and execute.

\subsection{Artifact check-list (meta-information)}

{\small
\begin{itemize}
  \item {\bf Algorithm: } \myCompilerName{}, MCTS
  \item {\bf Program: } Python
  \item {\bf Data set: } QECCs are in folder \verb|/qecc|
  \item {\bf Run-time environment: } Linux, Python
  \item {\bf Hardware: } x86, RAM $>16\ GiB$, Disk $>20\ GiB$
  \item {\bf Metrics: } Logical error rate
  \item {\bf Output: } Schedule of syndrome measurement
  \item {\bf Experiments: }Tables~\ref{tab:overall-error},~\ref{tab:cross-decoder}, Figures~\ref{fig:surface-result},~\ref{fig:bbcode},~\ref{fig:scaling},~\ref{fig:non-uniform}.
  \item {\bf Disk space required: $20\ GiB$}
  \item {\bf Time needed to prepare: $5mins$}
  \item {\bf Time needed to complete experiments: $<1hr$}
  \item {\bf Publicly available: accessed through \url{https://github.com/acasta-yhliu/asyndrome.git}}
  \item {\bf Code licenses: MIT}
  \item {\bf Data licenses: MIT}
  \item {\bf Archived: \doiurl}
\end{itemize}
}

\subsection{Description}

\subsubsection{How to access}

The artifact can be downloaded via public archive from Zenodo: \doiurl.

\subsubsection{Hardware dependencies}

Executing the artifact requires a Linux x86 server with RAM $>16\ GiB$ and disk $>20\ GiB$. A graphical interface is not required.

\subsubsection{Software dependencies}

Executing the artifact requires the user to have a Linux operating system and Python $>3.9$ installed. Other Python packages are listed within the artifact folder.

\subsubsection{Data sets}

The QECCs that are evaluated in our research are located in the \verb|/qecc| folder, in the following JSON format:
\begin{itemize}
    \item "family": name of the QECC, string.
    \item "n": number of physical qubits, integer.
    \item "k": number of logical qubits, integer.
    \item "d": code distance, integer.
    \item "logical\_xs": all logical $X$ operators, list of string.
    \item "logical\_zs": all logical $Z$ operators, list of string.
    \item "x\_stabilizers,z\_stabilizers": all stabilizers, list of string.
\end{itemize}

\subsection{Installation}

The user is recommended to use a Python virtual environment. After downloading the artifact and unzipping it, the user should navigate to the folder and set up the virtual environment through:
\begin{verbatim}
$ python3 -m venv venv
$ source venv/bin/activate
\end{verbatim}

Then install the necessary Python packages in the activated virtual environment:
\begin{verbatim}
(venv)$ pip3 install -r requirements.txt
\end{verbatim}

\subsection{Experiment workflow}

After installing the artifact, the user can use \verb|artifact.py| to reproduce results. The only argument it accepts is the asset the user wants to reproduce. For example, if you want to get Figure~\ref{fig:surface-result}, use:
\begin{verbatim}
(venv)$ python3 artifact.py figure12
\end{verbatim}

All available options are: \verb|table2|, \verb|table3|, \verb|figure12|, \verb|figure13|, \verb|figure14|, \verb|figure15|.

\subsection{Evaluation and expected results}

The program will output the corresponding figure/table to the \verb|/result/| folder. The figures are typeset as they are in the paper, and tables are in \LaTeX{} format. There may be numerical differences due to randomization.

\subsection{Experiment customization}

The user can use \verb|main.py| to customize their own schedule. Use \verb|python3 main.py --help| for more information and options.

\subsection{Methodology}

Submission, reviewing, and badging methodology:

\begin{itemize}
  \item \url{https://www.acm.org/publications/policies/artifact-review-and-badging-current}
  \item \url{https://cTuning.org/ae}
\end{itemize}